\documentstyle[preprint,prb,aps]{revtex}
\begin{document}
\title{The Generalized Gutzwiller Method for $n\geq 2$ Correlated Bands: First
Order Metal-Insulator Transitions}
\draft
\author{J. B\"{u}nemann and W. Weber}
\address{Institut f\"{u}r Physik, Universit\"{a}t Dortmund, \\
D-44221 Dortmund, Germany}
\date{\today}
\maketitle

\begin{abstract}
We have generalized the Gutzwiller method to the cases of $n\geq 2$
correlated bands and report studies on a degenerate two-band model with
Hund's rule type on-site interactions. At half band filling the
metal-insulator transitions are usually of first order.
\end{abstract}
\input{epsf}



% body of paper here
\section{Introduction}
The Gutzwiller variational method has been used since long for the study of
ground state properties of Hubbard-type models for correlated fermions. \cite
{c1} The one-band (one orbital) formulation has been extended in various
ways to treat, e.g. antiferromagnetism \cite{c2,c2b} or to include
additional, uncorrelated orbitals. \cite{c3c,c3,c3b} The Gutzwiller ansatz
has also been investigated analytically. \cite{c4} However, attempts to
extend the method to $n=2$ correlated orbitals per site have, sofar, led to
unsatisfactory results. \cite{c5}\\In this paper we generalize the
Gutzwiller method to cases of arbitrary numbers $n$ of correlated orbitals
per site. The extension makes it possible to study the ground state
properties of realistic models for magnetic 3d transition metals and
compounds. As a first application, we examine a two-band model with more
general on-site interactions and focus on the metal-insulator transition 
\cite{c6} at quarter and half band fillings.
\section{Generalization of the Gutzwiller Method}
For one band, the Gutzwiller variational wavefunction reads as 
\begin{equation}
\left| \Psi \right\rangle \equiv g^{\widehat{D}}\left| \Psi _0\right\rangle
=\prod_{s=1}^L\left[ 1-(1-g)\widehat{D}_s\right] \left| \Psi _0\right\rangle
\;,  \label{1}
\end{equation}
where $\left| \Psi _0\right\rangle $ is an uncorrelated wavefunction on a
lattice of $L$ sites and the operators $\widehat{D}_s\equiv \widehat{n}%
_{s\uparrow }\widehat{n}_{s\downarrow }$ measure the double occupancy of
sites $s$. The variational parameter $g$ will reduce the total number $%
D\equiv \overline{M}_{12}$ of doubly occupied sites and may vary between $g=1
$ ($\overline{M}_{12}=M_1M_2/L$) and $g=0$ ($\overline{M}_{12}=0$). Here $%
M_1(M_2)$ are the gross numbers of spin up(down) sites. The norm of (\ref{1}%
) is approximated by \cite{c7} 
\begin{equation}
\left\langle \Psi \mid \Psi \right\rangle
=P_1P_2\sum_Dg^{2D}N_D(L,M_1,M_2)\;,  \label{4}
\end{equation}
with the probabilities $P_k=M_k!(L-M_k)!/L!$ and the combinatorial weight
factors 
\begin{equation}
N_D\equiv \frac{L!}{\overline{M}_0!\overline{M}_1!\overline{M}_2!\overline{M}%
_{12}!}\;.  \label{5}
\end{equation}
Here we have introduced net numbers of occupied sites $\overline{M_k}=M_k-%
\overline{M}_{12}$ and of empty sites $\overline{M}_0=\left( L-\overline{M}%
_1-\overline{M}_2-\overline{M}_{12}\right) $. In the thermodynamic limit the
sum of eq. (\ref{4}) is replaced by its largest term, leading to the
relation 
\begin{equation}
g^2=\overline{m}_0\overline{m}_{12}/(\overline{m}_1\overline{m}_2)
\label{10}
\end{equation}
with $\overline{m}_0=\overline{M}_0/L$, etc. Also the expectation values 
\begin{equation}
\left\langle \Psi \left| \widehat{c}_{s\sigma }^{+}\widehat{c}_{t\sigma
}\right| \Psi \right\rangle /\left\langle \Psi \mid \Psi \right\rangle
=q_\sigma \left\langle \widehat{c}_{s\sigma }^{+}\widehat{c}_{t\sigma
}\right\rangle _0,\quad s\neq t,  \label{11}
\end{equation}
are found by combinatorics. Here the ``loss factors'' 
\begin{equation}
q_k\equiv \frac 1{m_k(1-m_k)}\left[ \sqrt{\overline{m}_k\overline{m}_0}+%
\sqrt{\overline{m}_{kl}\overline{m}_l}\,\right] ^2  \label{12}
\end{equation}
with $k,l=1,2$ and $k\neq l$ represent the step sizes of the momentum
distribution functions at the Fermi level ($q_k=1$ for $g=1$).

When we extend the Gutzwiller method to cases of $n\geq 2$ correlated
orbitals per site, the number of different site occupancies $J_n=4^n$ will
increase enormously. For $n=2$ and using the notation $a\!\!\uparrow \!\!%
\widehat{=}1,$ $a\!\!\downarrow \!\!\widehat{=}2,$ $b\!\!\uparrow \!\!%
\widehat{=}3,$ $b\!\!\downarrow \!\!\widehat{=}4$ the $16$ occupancies are:
i) empty ($\overline{M}_0$); ii) four single ($\overline{M}_k~,k=1,...,4$);
iii) six double ($\overline{M}_{kl}~,k,l=1,...,4;k\neq l)$; iv) four triple (%
$\overline{M}_{klp})$ and v) quadruple ($\overline{M}_{1234}$). In general,
the $J_n$ occupancies are labeled by subscripts $I_j$, which are assigned to
the multiple orbital-spin subscripts by 
\begin{equation}
\{I_1,I_2,...,I_{2n+1},I_{2n+2},...,I_{J_n}\}=\{(0),(1),...,(2\cdot
n),(12),...,(k...p)\}\;.  \label{12b}
\end{equation}
The net and gross occupancies are measured by appropriate operators, e.g., 
\begin{equation}
\widehat{\overline{M}}_{123}\equiv \sum_s\widehat{n}_{s\uparrow }^{(a)}%
\widehat{n}_{s\downarrow }^{(a)}\widehat{n}_{s\uparrow }^{(b)}(1-\widehat{n}%
_{s\downarrow }^{(b)})=\widehat{M}_{123}-\widehat{M}_{1234,}  \label{13}
\end{equation}
There are now $K_n=J_n-(2n+1)$ multiple occupancies so that the generalized
Gutzwiller operator reads as 
\begin{equation}
g\widehat{^D}\rightarrow \prod_{j=2n+2}^{J_n}\left( g_{I_j}\right) ^{%
\widehat{\overline{M}}_{I_j}}\;.  \label{15}
\end{equation}
Note, that in contrast to ref. \onlinecite{c5}, we found it essential to include 
\underline{all} multiple occupancies and to use net multiple occupancy
operators. The norm is given by 
\begin{equation}
\left\langle \Psi \mid \Psi \right\rangle =\prod_{k=1}^nP_k\sum_{\widetilde{D%
}}N_{\widetilde{D}}\prod_{j=2n+2}^{J_n}\left( g_{I_j}\right) ^{2\overline{M}%
_{I_j}}  \label{16}
\end{equation}
with the $P_k$ factors of eq. (\ref{4}), and 
\begin{equation}
N_{\widetilde{D}}\equiv L!\left[ \prod_{j=1}^{J_n}\overline{M}_{I_j}!\right]
^{-1}  \label{17}
\end{equation}
The sum of eq. (\ref{16}) includes all sets of multiple occupancy
configurations $\widetilde{D}=\left\{ \overline{M}_{I_j}\right\} $ with $%
j=\left( 2n+2\right) ,...,J_n$. Replacing the sum by its largest term leads
to the $K_n$ relations 
\begin{equation}
g_{I_j}^2=\prod_{i=1}^{J_n}\left( \overline{M}_{I_i}\right) ^{\chi
_{ij}^{}}\;,\quad \text{with }\chi _{ij}^{}=\frac{\partial \overline{M}_{I_i}%
}{\partial \overline{M}_{I_j}}\;.  \label{19}
\end{equation}
Keeping the gross numbers $m_k$ fixed, we get 
\begin{eqnarray}
g_{kl}^2 &=&\overline{m}_0\overline{m}_{kl}/\left( \overline{m}_k\overline{m}%
_l\right)  \nonumber\\
&&.....  \nonumber \\
g_{k..p}^2 &=&\left( \overline{m}_0\right) ^{\left( \gamma -1\right) }%
\overline{m}_{k..p}/\left( \overline{m}_k...\overline{m}_p\right) \;, 
\label{23}
\end{eqnarray}
where $\gamma $ is the number of subscripts in $\overline{m}_{k..q}$. For
each of the generalized ``loss factors'' $q_{kl}$ the evaluation of $%
4^{2n-1}$ terms is required. Nevertheless, the resulting equations for the $%
q_{kl}$ factors look remarkably simple: 
\begin{mathletters}
\label{25}
\begin{eqnarray}
q_{kk} &=&\frac 1{m_k(1-m_k)}[\sqrt{\overline{m}_k\overline{m}_0}%
+\sum_l{}^{^{\prime }}\sqrt{\overline{m}_{kl}\overline{m}_l}  \nonumber \\
&&+\sum_{l,p}{}^{^{\prime \prime }}\sqrt{\overline{m}_{klp}\overline{m}_{lp}}%
+\sum_{l,p,q}{}^{^{\prime \prime \prime }}\sqrt{\overline{m}_{klpq}\overline{%
m}_{lpq}}+...]^2  \label{25a} \\
q_{kl}^2 &=&q_{kk}q_{ll}  \label{25b}
\end{eqnarray}
\end{mathletters}
Here the summation primes exclude $k=l$, etc. Obviously eqs. (\ref{25}) are
a straightforward generalization of eq. (\ref{12}) and of results obtained
in ref. \onlinecite{c3c}. Simpler forms of the $q_{kk}$ equations are obtained,
when certain multiple occupancies are put zero, as was assumed in ref. \onlinecite{c8}. Recently, eqs. (\ref{25}) have also been found by T. Okabe. \cite{c8b}

Using the generalized Gutzwiller wavefunction, we can now investigate
extensions of the Hubbard model for arbitrary numbers of orbitals $\alpha
,\beta $: 
\begin{eqnarray}
\widehat{H} &=&\sum_{\alpha,\beta,s,t,\sigma} T_{st}^{\alpha \beta }\widehat{%
\alpha }_{s\sigma }^{+}\widehat{\beta }_{t\sigma }+\sum_
{\alpha,\beta,s,\sigma,\sigma ^{\prime }} \!\!\!\!^{^{\prime }}U^{\alpha
\beta }\widehat{n}_{s\sigma }^\alpha \widehat{n}_{s\sigma ^{\prime }}^\beta 
\nonumber \\
&&+\sum_{ \alpha ,\beta ,s ,\sigma ,\sigma ^{\prime } } \!\!\!\!^{^{\prime
}}J^{\alpha \beta }\widehat{\alpha }_{s\sigma }^{+}\widehat{\beta }_{s\sigma
^{\prime }}^{+}\widehat{\alpha }_{s\sigma ^{\prime }}\widehat{\beta }%
_{s\sigma }\;,  \label{27}
\end{eqnarray}
which include, apart from general hopping terms, the on-site interaction $U$%
, both orbital-diagonal $\left( U^{\alpha \alpha }\right) $ and off-diagonal 
$\left( U^{\alpha \beta },\alpha \neq \beta \right) $, and the on-site or
Hund's rule exchange $J^{\alpha \beta }$. The primes exclude non-physical
orbital-spin combinations in the respective $U$ and $J$ summations. For the
expectation values of the on-site exchange only the terms 
\begin{equation}
-\sum_{\alpha ,\beta ,\sigma }\!\!^{^{\prime }}J^{\alpha \beta }\widehat{n}%
_{s\sigma }^\alpha \widehat{n}_{s\sigma }^\beta  \label{29}
\end{equation}
survive, i.e., all spin fluctuations are ignored.

By assuming $J^{\alpha \beta }<U^{\alpha \beta }<U^{\alpha \alpha }$, it is
possible to achieve the filling of incomplete $d$ shells in the atomic limit
according to Hund's rules. Thus, the Hamiltonian of eq. (\ref{27})
represents realistic models for the valence electron structure of both 3d
transition metal elements and of magnetic 3d oxides and halides, provided
that a sufficient number of orbitals and sites are included. The $N-$%
particle wavefunctions $\left| \Psi \right\rangle ,\left| \Psi
_0\right\rangle $ can be chosen to represent para-, ferro-,\- or
antiferromagnetic states, and in principle, even more complicated cases,
such as orbital ordering. Also, ligand orbitals or $s,\,p$ type orbitals on
the $d$ sites may be incorporated. However, any additional complexity of the
wavefunctions $\left| \Psi _0\right\rangle $ leads to further variational
parameters in $\left| \Psi _0\right\rangle $ representing effective orbital
energy shifts or effective hoppings, etc. \cite{c3c} The problem of orbital
density depletion caused by the reduction of multiple occupancies can be
treated in analogy to ref. \onlinecite{c3b}.
\section{Two-Band Model with orbital degeneracy}
In the following, we investigate the $n=2$ Hamiltonian, i.e., we assume one
atomic site on a simple cubic lattice with two, degenerate $d(e_g)$
orbitals, and we study only paramagnetic solutions. The single particle
bands are constructed using realistic $1NN$ and $2NN$ hopping matrix
elements $T_{dd\sigma }(1NN)=1eV,$ $T_{dd\sigma }(2NN)=0.25eV$, and $%
T_{dd\delta }:T_{dd\pi }:T_{dd\sigma }=0.1:-0.3:1$. There are three
interaction parameters $U^{\alpha \alpha }\equiv U,\;U^{\alpha \beta }\equiv
U^{\prime }$ and $J^{\alpha \beta }\equiv J$, which, in the limit of
vanishing configuration interaction are related by \cite{c10} 
\begin{equation}
J=\frac 12(U-U^{\prime })\,.  \label{29b}
\end{equation}
For the paramagnetic case we have $m_k\equiv m,\;\overline{m}_{klp}\equiv
t,\;m_{1234}\equiv f,$\ $\overline{m}_{12}=\overline{m}_{34}\equiv d_d,\;%
\overline{m}_{14}=\overline{m}_{23}\equiv d_o,\;$and $\overline{m}_{13}=%
\overline{m}_{24}\equiv d_t$. All $q_{kl}$ are equal 
\begin{eqnarray}
q &=&\frac 1{m(1-m)}[(\sqrt{d_d}+\sqrt{d_o}+\sqrt{d_t})(\sqrt{\overline{m}}+%
\sqrt{t})  \nonumber \\
&&+\sqrt{\overline{m}_0\overline{m}}+\sqrt{tf}]^2\,.  \label{30}
\end{eqnarray}
The energy function for $e/a\equiv 4m$ (number of electrons per atom) is 
\begin{eqnarray}
E &=&2q\overline{\epsilon }(m)+2Ud_d+2U^{\prime }d_o+2(U^{\prime }-J)d_t
\nonumber \\
&&+(2U+4U^{\prime }-2J)(2t+f)\;.  \label{31}
\end{eqnarray}
Here $2\overline{\epsilon }(m)$ is the kinetic energy of the uncorrelated
case$.$ $E$ has been minimized numerically with respect to the five
variational parameters $d_d,d_o,d_t,t,f$ and has been studied at several $m$
 values and for various sets of interaction parameters $U$ and $u^{\prime
}\equiv U^{\prime }/U$, keeping $2J=U-U^{\prime }$ according to eq. (\ref
{29b}).

There are two $m$ regions of interest. The first is around $m=0.25$ (or $%
m=0.75$), the other around $m=0.5$. For these integer fillings $(e/a=1,2,3)$
we observe metal-insulator (MI) transitions in the $(U,u^{\prime })-$plane.
However, there are big differences between the two cases, mainly concerning
the shapes of the phase diagrams in the $(U,u^{\prime })-$plane and the
order of the MI transitions.

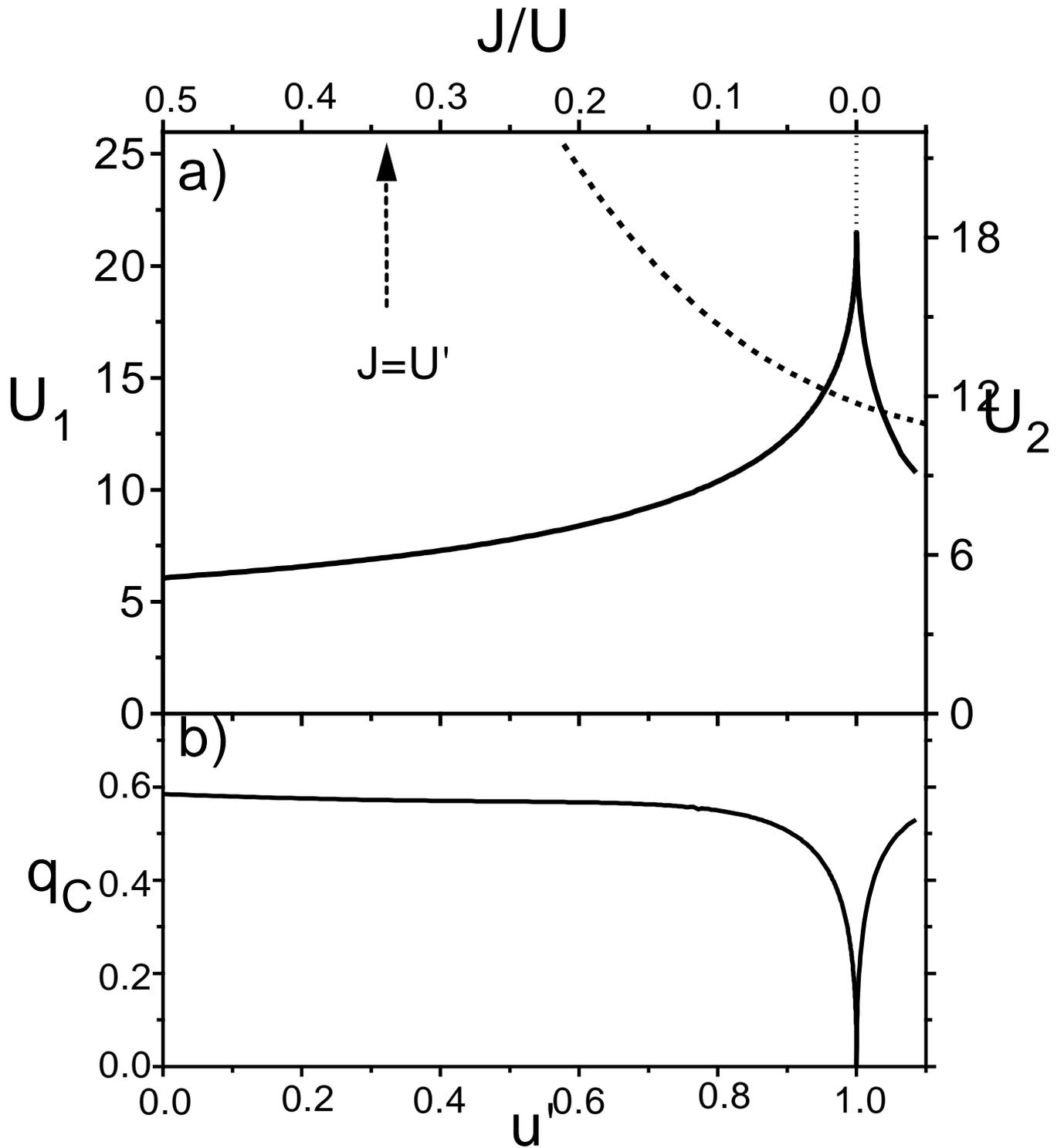
\begin{figure}
\caption{a) Phase diagrams in the $(U,u^{\prime })-$plane of metallic and
insulating phases for $m=0.25$ (dashed line, LH scale $U_1$) and for $m=0.5$
(solid line, RH scale $U_2$). The lines separate the metallic phases (small $%
U$) from the insulating phases (large $U$). For $m=0.5$ the insulating phase
is a triplet state $(d_t=0.5)$ for $u^{\prime }<1$, and a singlet state $%
(d_d=0.5)$ for $u^{\prime }>1$. b) Loss factor $q_C$ at the MI transition
points $(U_C,u_C^{\prime })$.}
\label{f1}
\end{figure}

For $m=0.25$, the MI phase diagram is shown in Fig. \ref{f1}a. For $u^{\prime }=1$, we find $U_C\approx 13.8$ in agreement with ref.\onlinecite{c8}. For $u^{\prime }>1/3$ $(U^{\prime }>J)$ the system will avoid all multiple occupancies in the
limit $U\rightarrow \infty $, so that there is always a MI transition. This
transition is of second order in $q$, very similar to that of the one-band
model. \cite{c6} Near any critical values $(U_C,u_C^{\prime })$, $q$ behaves
like $q\approx q_0(1-U/U_C)$, when we approach $U_C$ from the metallic side,
i.e.; the effective mass $m^{*}=m_e/q$ diverges with $(1-U/U_C)^{-1}$. \cite
{c6,c7} For $u^{\prime }\leq 1/3$ $(U^{\prime }\leq J)$ the system is
always metallic.

In contrast, the MI transition for $m=0.5$ is very different (see Fig. \ref{f1}a).
The transition is generally of first order in $q$. This means that the $q$
values change discontinuously at critical values ($U_C,u_C^{\prime })$ from
a metal (typically $q\approx 0.6$) to an insulator ($q=0$) (see Fig. \ref{f1}b).
Only at the point $u^{\prime }=1\,($i.e. $U_C=U_C^{\prime })$ the MI
transition is second order in $q$. Note that apart from the point $u^{\prime
}=1$ the ground state energy $E$ varies at the transition in such a way that
its first derivatives with respect to all relevant system parameters changes
discontinuously. In particular we observe a finite discontinuity of the
chemical potential $\mu (m)=\partial E(m)/\partial m$ at $m=0.5$ for all $%
u^{\prime }<1$ if $U$ exceed $U_C$ by any infinitesimal amount.

In the insulating region we have $t=f=0$, and because of particle hole
symmetry, $\overline{m}_0=\overline{m}=0.$ For $u^{\prime }<1$ only the spin
triplet occupancy $d_t$ prevails: $d_t=0.5,\,d_d=d_o=0$. The case $u^{\prime
}=1$ is special, as there is no preference for one of the three double
occupancies. We believe, because of this kind of frustration the metallic
phase extends so far out near $u^{\prime }\lesssim 1$.

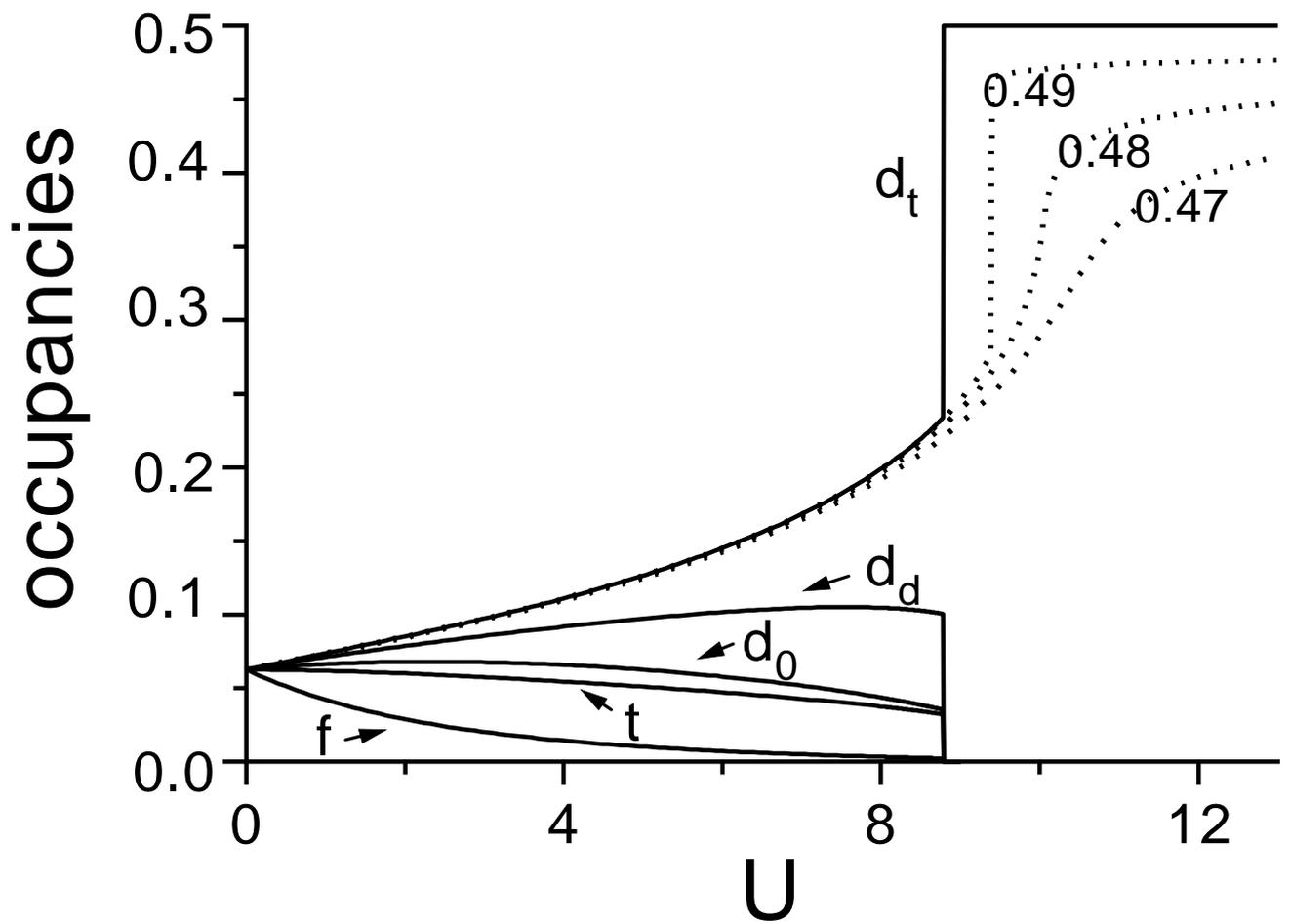
\begin{figure}
\caption{Double $(d_t,d_d,d_o)$, triple $(t)$ and quadruple $(f)$
occupancies for $m=0.5$ $(e/a=2)$ as a function of $U$, with $u^{\prime
}=0.8 $ (solid lines). Also shown are $d_t(U)$ curves for $%
m=0.49,\;0.48,\;0.47$, respectively, indicating the ``good-to-bad'' metal
transition away from $m=0.5$.}
\label{f2}
\end{figure}

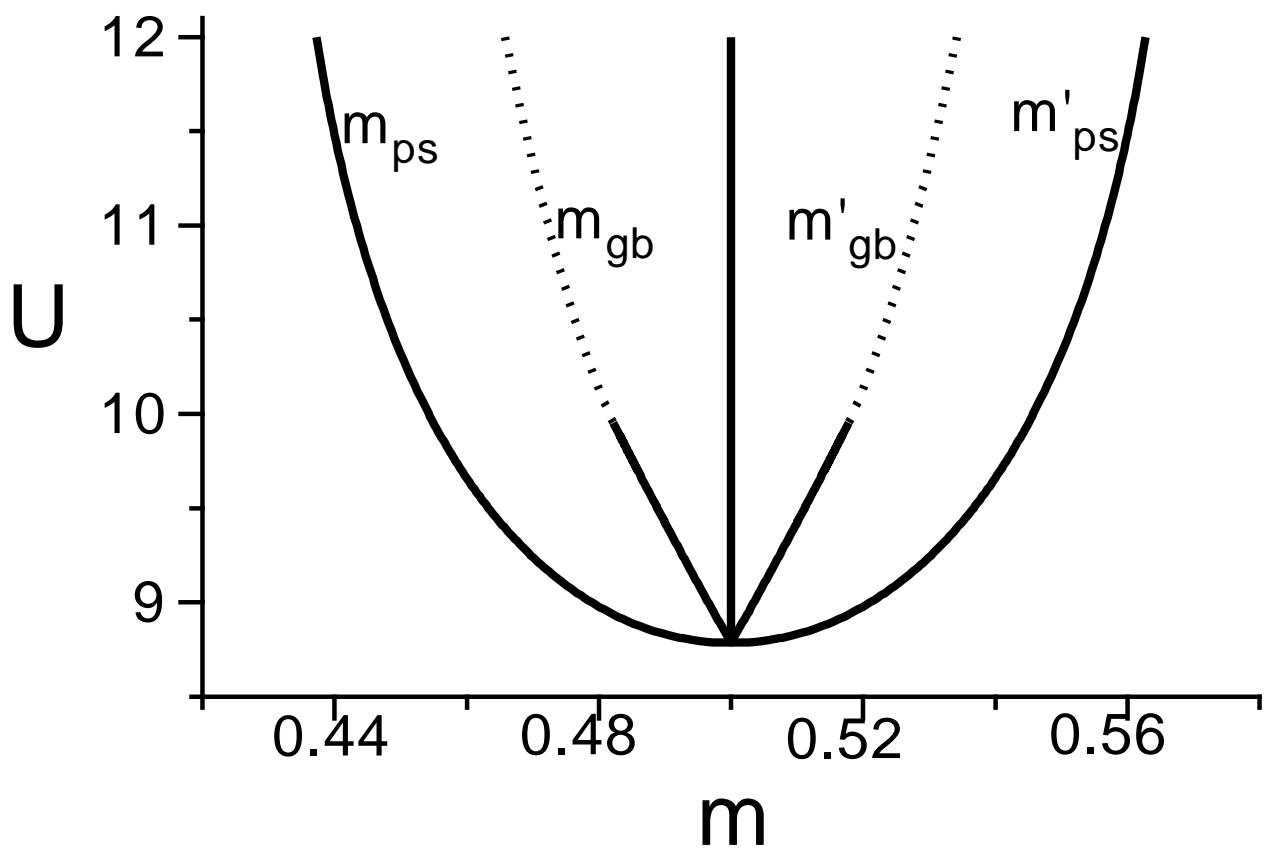
\begin{figure}
\caption{Phase diagrams in the $(U,m)-$plane, with $u^{\prime
}=0.8 $: Lines dividing the ``good
metal'' from the ``bad metal'' phases are indicated by $m_{gb}$, lines
dividing the homogeneous from the phase separated regions are denoted by $%
m_{ps}$. For the part of $m_{gb}$ given by dashed lines, the discontinuity
in $q(U,u^{\prime })$ has disappeared; here inflection points are used to
determine the phase boundary.}
\label{f3}
\end{figure}

In the metallic regime all multiple occupancies are nonzero (see Fig. \ref{f2}).
For $u^{\prime }<1$, $d_t$ increases with $U$, typically up to $d_t\approx
0.25$. Then, at $U_C$, $d_t$ jumps up to $0.5$, representing a
low-to-high-spin transition. For certain band filling values close to $m=0.5$%
, the first order step is still present in the $q(U,u^{\prime })$ and $%
d_t(U,u^{\prime })$ curves. Further away from $m=0.5$, the curves smoothen,
yet still vary rapidly in the range of critical parameters $(U_C,u_C^{\prime
})$ as shown in Fig. \ref{f2}. A typical phase diagram in the $(U,u^{\prime })-$%
plane of this ``good-to-bad''-metal or low-to-high-spin transition is shown
in Fig. \ref{f3}.

Note that for our evaluation a homogenous phase was assumed. However, in the
vicinity of the low-to-high-spin transition, the homogeneous phase is
unstable against phase separation. As a consequence there exists a surface $%
m_{ps}(U,u^{\prime })$ in the space of $(U,u^{\prime },m)$ which divides the
homogeneous from the separated region (see Fig. \ref{f3}). At any point $%
(U,u^{\prime },m)$ with $m_{ps}<m<0.5$ (or $m_{ps}^{\prime }>m>0.5)$, the
system breaks up into one part with density $m_1=0.5$ and another one with $%
m_2=m_{ps}(m_{ps}^{\prime })$. It is an open question, however, whether the
phase separation is an artefact of Hubbard-type models, because of, e.g.,
the neglect of long range Coulomb interaction.

The first order behavior of the MI transition for $m=0.5$ is caused by the
competition of two different energy minima, now possible due to of the much
larger space of variational parameters. We expect that the first order
nature will survive when we include antiferromagnetic states in the
minimization. First order, low-to-high spin transitions are also observed
for two-band, itinerant ferromagnetic states. \cite{c9} In fact, we suspect
that first order changes in the electronic structure of incomplete 3d shells
may drive many phase transitions of magnetic 3d compounds. It should be
interesting to reexamine the MI phase transitions observed e.g. for $%
Ni\,S,\;Ni\,J_2$ or $RE\,Ni\,O_3$\cite{c11,c12,c13} with respect to their
first order behavior.

Very recently, a dynamical mean field study about the MI transition in a degenerate (n=2) Hubbard model has been presented. \cite{c14} The authors only consider the cases $J=0$ and/or $m\leq0.25$ (in our notation). In aggreement with our results, they find a qualitative feature, which is very similar to the $n=1$ case. Nevertheless, their more general remark that the nature of the Mott transition is not qualitatively changed by orbital degeneracy is not supported by our results, concerning the cases $J>0,\,m\approx0.5$.
\section{Concluding remarks}
In conclusion, we have extended the Gutzwiller method to arbitrary numbers $%
n\geq 2$ of correlated orbitals per site and have reported studies on the
simplest $n=2$ model. We find a wealth of new features, not present in the $%
n=1$ models, such as first order metal-insulator transitions or low-to-high
spin transitions. Numerical studies appear to be feasible even for $n=5$
realistic models of 3d transition metals and compounds.

Acknowledgments: This work has been partly supported by the European Union
Human Capital and Mobility program, Project No. CHRX-CT 93-0332

% now the references. delete or change fake bibitem. delete next three
%   lines and directly read in your .bbl file if you use bibtex.

% \begin{figure}
% \caption{}
% \label{}
% \end{figure}

%  fig. 1

% fig. 2

%  fig. 3

% \begin{table}
% \caption{}
% \label{}
% \end{tabular}
% \end{table}

\end{document}